# Title Page

**First author and Corresponding author:**

Seyedamiryousef Hosseini Goki

Department of Computer Science

University of Victoria, Victoria, BC, Canada

yousofhosseini9877@uvic.ca

**Second author:**

Dr. Mahdieh Ghazvini

Department of Computer Engineering

Shahid Bahonar University of Kerman, Kerman, Iran
mghazvini@uk.ac.ir

**Third author:**

Sajad Hamzenejadi

Department of Industrial and Information Engineering

Politecnico di Milano, Milan, Italy

sajad.hamzenejadi@mail.polimi.it

**Conflict of Interest:** The authors declare that they have no conflict of interest.

# A Wavelet Transform Based Scheme to Extract Speech Pitch and Formant Frequencies

*Abstract*— Pitch and formant frequencies are important features in speech processing applications. The period of the vocal cord's output for vowels is known as the pitch or the fundamental frequency, and formant frequencies are essentially resonance frequencies of the vocal tract. These features vary among different persons and even words, but they are within a certain frequency range. In practice, just the first three formants are enough for the most of speech processing. Feature extraction and classification are the main components of each speech recognition system. In this article, two wavelet based approaches are proposed to extract the mentioned features with help of the filter bank idea. By comparing the results of the presented feature extraction methods on several speech signals, it was found out that the wavelet transform has a good accuracy compared to the cepstrum method and it has no sensitivity to noise. In addition, several fuzzy based classification techniques for speech processing are reviewed.

## I. INTRODUCTION

Speech signal is a type of periodic waveforms in the time domain, whose frequency is variable with time. Therefore, it is necessary to pay attention to both the information obtained from the signal in time and frequency domains. Signals that do not change their frequency over time are called stationary signals. Obviously, the speech signal is a non-stationary signal. The Fourier transform gives us the spectral content of the signal, but in terms of time it only pays attention to the whole signal over the time and it has no focus on the time domain, so it's not suitable for non-stationary signals that vary over the time. The wavelet transform solves this issue by its concurrent accuracy in both frequency and time domain. In other words, the wavelet transform gives the time-frequency information simultaneously.

Discrete Wavelet Transform (DWT) is much simpler than Continuous Wavelet Transform (CWT), since it has less number of calculations. In the case of discrete wavelet transform, we use different frequency cutoff filters at various scales to analyze an input signal. For example, a signal is conducted from low pass and high pass filters to analyze low and high frequencies, respectively. After each passage of the filters, it takes a drop in the sampling rate. Because after passing the filter the frequency of the signal is halved, and according to the Nyquist rate, the sampling rate can be reduced by half. Signal decomposition by wavelet transform produces a bunch of band pass signals with different frequencies and different precisions. This decomposition is based on the multi resolution characteristic of wavelet transform that is designed for high time accuracy and poor frequency accuracy at high frequencies signals and high frequency accuracy and poor time accuracy at low frequencies signals. Since the formant and pitch frequencies are roughly in different frequency bands, their effects can be separated from each other by the multi resolution feature of wavelet transform. Therefore, in order to determine the formants since the formant frequencies are higher than the pitch frequencies, after the conversion, the corresponding coefficients with low frequencies and pitch frequencies reduced to zero and then the signal is analyzed. In other words, only lower-scale coefficients are used to analyze the signal and the formant frequencies are extracted from the reconstructed signal spectrum. Also with this idea, we can eliminate the coefficients of the high frequency component of the signal and reconstruct the signal only by using the frequency coefficients that correspond to the low-frequency components of the signal. This property is used to determine the speech pitch frequency. Therefore, wavelet transform is a suitable procedure for determining the speech pitch and formant frequencies. The primary version of the current paper has been presented in [1]. The parameters determined by this method have high accuracy and can be used in speech processing applications such as speech recognition and production systems, speaker authentication systems, speaker identification, speech compression and coding.

In this article, we proposed two novel approaches to extract pitch and formant frequencies based on wavelet transform regard to the filter bank idea. Speech feature estimation plays a vital role in executing most of processes that are executed on speech signal. But some problems such as the flexibility of human vocal tract, co-articulations, speaker pronunciation, prosody and emotional states make this estimation a tough work. Pitch and formant frequencies as significant features in speech processing applications are commonly used by classifiers for several purposes such as speech and speaker recognition, speech segmentation, speech gender recognition, and speech emotion recognition. Fuzzy logic is one of the most efficient classifiers for speech processing. In the following, the usage of this classifier in speech processing applications is discussed.

A list of abbreviations and acronyms used throughout the paper is given in Table 1. The rest of the paper is organized as follows. Section II introduced the wavelet transform, speech organs and speech recognition systems. Section III describes feature extraction in speech processing. Some applications of fuzzy classifier in speech analysis is introduced in Section IV. The proposed method is presented in Section V and Section VI demonstrates the performance evaluation. Finally, Section VII concludes the paper.

Table 1. List of acronyms and abbreviation

| | |
|---|---|
| **ACO** | Ant Colony Optimization |
| **AMDF** | Average Magnitude Difference Function |
| **ANFIS** | Adaptive Neuro-Fuzzy Inference System |
| **CWT** | Continuous Wavelet Transform |
| **DFCC** | Diatonic Frequency Cepstral Coefficients |
| **DTW** | Dynamic Time Wrapping |
| **DWT** | Discrete Wavelet Transform |
| **EE** | Energy Entropy |
| **FL** | Fuzzy Logic |
| **FoGs** | **formant-gaps** |
| **GA** | Genetic Algorithm |
| **HP** | High Pass |
| **LP** | Low Pass |
| **LPC** | Linear Predictive Coding |
| **MFCC** | Mel Frequency Cepstrum Coefficients |
| **MVML** | Multi Views Multi Learners |
| **MVSL** | Multi-view Semantic Learning |
| **NN** | Neural Network |
| **PSO** | Particle Swarm Optimization |
| **SER** | Speech Emotion Recognition |
| **STE** | Short Time Energy |
| **SVM** | Support Vector Machine |
| **ZCR** | Zero Crossing Rate |

II. LITERATURE REVIEW

A fast review on speech organs and wavelet transform as well as speech recognition is presented in this section.

*A. Speech Organs*

Features that determine voice characteristics are related to the size of vocal cavity, speech organs, the location and the way that the speech signal is made, fundamental frequency (pitch) and sub-waves. The rapid opening and closing of the vocal cords causes an important physical phenomenon called the vowel. When the vocal cords come together, they block the path of the air to the outside. If this obstruction is not such that its opening requires a lot of pressure, with a slight lung air pressure that it is concentrated under the vocal folds, the vocal folds are slightly opened and air will come out. At this time, due to the lower air pressure of the lungs, the muscle strength again closes the path of air to the outside by shuffling the vocal folds. If these openings and closings happen very fast and periodically (more than 16 times per second), the vowels will be created. Men's vocal folds are thicker and longer than women's vocal folds because the amount of vibration depends on the thickness and length of the vibrating body, hence the fundamental frequency of men's voice is usually less than the women's one. In other word, the muscular structure of the vocal folds is such that their length and thickness can be intentionally changed, with such a mechanism, the quality of sound is altered in terms of pitch frequency. Albeit, these pitch changes in humans have a limited domain. Usually this frequency limitation in the voice of a normal person is between 70 and 1000 cycles per second, although low frequencies belong to the sound of men and high frequencies are for women's voices. In the texture of speech, the frequency of vocal cords is changing, and other vocal elements created by the acoustic vocal cavity, larynx, (such as pharynx, mouth, nasal, etc.) are combined with the fundamental frequency of vocal cords (pitch). Each specific frequency created by the aforementioned cavities is called formant. In other words, the frequency of the vocal tract amplification is called "formant". Generally, the individual voice quality for each person depends on several factors, one of which is the fundamental sound frequency or the speech pitch.

*B. Wavelet Transform*

Wavelets are building blocks for analyzing functions and common signals. In other words, any common function can be written as a linear combination of wavelets. In addition, by using wavelet coefficients, you can get a more compact view of the signal's initial representation. In wavelet transform, basis functions are not particular and unlike Fourier transformation, they are not limited to the stationary signals. Hence, for non-stationary signals, wavelet transformation can also be used. In addition, since wavelet analysis is similar to the analysis performed in the human ear, this transformation is suitable for speech signal processing. The wavelet expansion extends the single-dimensional signal to a double-dimensional array of coefficients. This double-dimensional display lets you focus the signal both in time and frequency.

A wavelet transform as a time-frequency transform with its multi resolution analysis is capable for displaying the signal's characteristics in both time and frequency domains (scales). Thus, it is useful for analyzing non-stationary signals. Since the speech signal is non-stationary and time-varying, it is obvious that for analysis of this signal, the methods for analyzing the non-stationary signals such as wavelet transformation should be used. The analysis of the speech signal with the wavelet transform is similar to the analysis performed in the human ear, so it is in accordance with the auditory sensation (perception). Wavelet decomposition methods provide a new group of basic functions that have better time-frequency differentiation properties. The analysis of the signal by using a series of samples with different precisions is referred to multi resolution analysis. Accuracy is in fact an expression of the frequency's contents of the signal. In reality, obtaining a signal with multi resolution accuracy is equivalent to passing through the signal of low pass filters, because each signal component is filtered (high frequency components). Typically, the signals which are being processed have information that appears as a sudden jump.

*C. Fundamental of Speech Recognition systems*

Each Speech (Speaker, Gender, Emotion…) Recognition system has two basic steps: feature extraction and classification. Feature extraction methods play significant role in these recognition systems. These methods are divided to temporal domain (parametric) and frequency domain (non-parametric) approaches. An example of parametric temporal domain approaches is the linear prediction coding (LPC), which is introduced to attune the resonant model of the human's vocal tract. LPC estimates the formants and remove their effects from the speech signal. It is one of the useful speech analysis schemes for speech encoding at a low bit rate. Besides it gives good estimations of speech parameters. The second category introduces the non-parametric frequency domain approaches such as the Mel-Frequency Cepstral Coefficients (MFCC) [2, 3]. The classification step has also a critical role in the speech recognition systems. Nowadays, many classification methods are utilized to distinguish speech signals. The recognition approaches may be classified into several categories: Neural Network based recognition, Fuzzy Logic based recognition, Wavelet Transform based recognition, Optimization Algorithms based recognition, Dynamic Time Wrapping algorithm based recognition, Sub-band based recognition and etc. [3]. The general architecture of a (Speech, Speaker, Emotion, Gender, etc.) recognition system is shown in Fig.1.

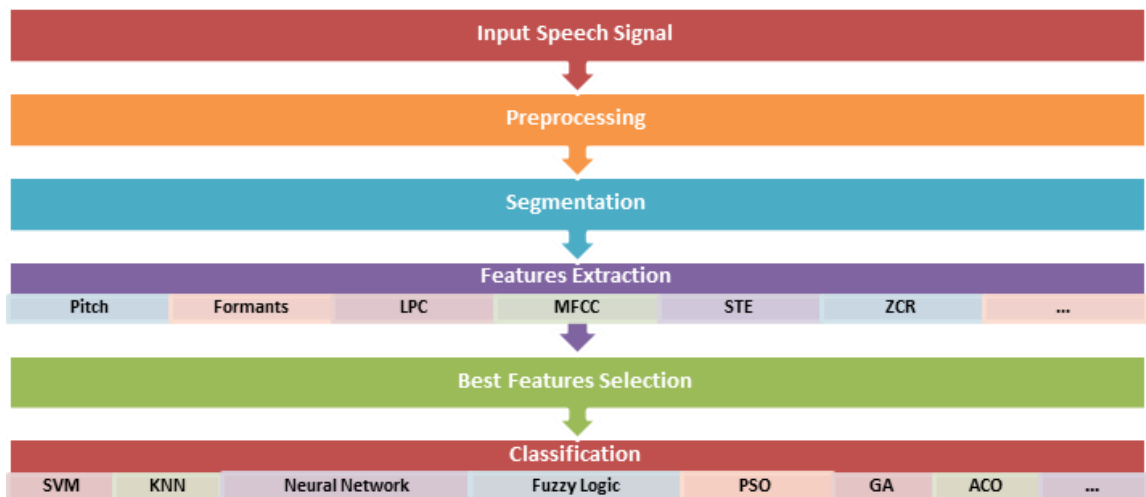

Fig1. The General Architecture of a (Speech, Speaker, Emotion, Gender, etc.) Recognition System

## III. FEATURES EXTRACTION

The estimation of speech features is an important step in performing most of the processes that are performed on the speech signal. But with regard to issues such as the flexibility of human vocal tract, co-articulations, speaker pronunciation, prosody and emotional states make the speech features estimation a difficult task. Pitch and formant frequencies are important features in speech processing applications [4]. Generally, the methods of speech pitch extraction are categorized as follows: time domain methods, frequency domain methods, and time-frequency domain methods. Among the methods of the time domain the autocorrelation function and the average magnitude difference function can be pointed out. In the frequency domain, harmonic product spectrum and cepstrum are commonly used. In the time-frequency domain wavelet transformation is also very important and useful. In another perspective, speech pitch extraction algorithms are categorized into two groups of event detection and non-event detection algorithms. The event detection algorithms are based on autocorrelation function. The defects of the event detection algorithms based on the autocorrelation function are that these algorithms estimate the pitch period in exactly one vowel, and because of this, the efficiency of these algorithms is reduced in places where pitch period is non-stationary. The non-event detection algorithms extract the pitch period with a direct method, so it is easier to calculate compared to event detection algorithms on a windowed speech signal.

In non-event detection methods, the average of pitch period on a piece of the speech signal which is obtained from a fixedlength window is calculated by one of the other methods, such as the calculation of the cepstrum and AMDF. These methods are not sensitive to changes in the non-stationary pitch period. Therefore they are not suitable for a wide range of speakers. Wavelet transform is appropriate due to its multi resolution property for analyzing the non-stationary speech signal. Also wavelet-based methods have a better resistance to noise.

One of the benefits of wavelet transformation is the processing of the speech signal without the assumption of being a stationary or non-stationary in the window. So, it is suitable for a wide range of pitch periods and it is capable for detecting the beginning of the pitch and the numbers of the pitch periods that are in one segment. Since the wavelet transform is computed for a few limited scales, the complexity of the calculation is not very high. So, it has a better performance than other methods such as autocorrelation and cepstrum. Formant frequencies are the escalation frequencies of vocal tract. In general, it can be said that the formant is a group of harmonic frequencies that are combined with the basic frequency, and determines the tone of voice. Formants are divided into two groups: high formants and low formants. The first group has higher frequencies and the second group includes low frequencies. Due to phonological experiments, it has been shown that the two main formants, high and low, which are characterized by the signs F1 and F2 are the most important formants of vowels. These two formants play a vital role in creating vowels or tones. For each person and even each word, these frequencies are different but they are roughly around certain frequencies. In reality, for the purpose of coding and other processes, the calculation of three formants is sufficient. The most common method for estimating formants is the use of a cepstrum.

Extraction of formant frequency in the processing, recognition and production of speech is important. There are several methods to extract these frequencies such as a cepstrum and linear prediction coding. Yan et al. [5] used pitch and formant frequencies as the feature set of each voiced speech segment and proposed a DTW based method to find out copy-move forgeries in speech recording. A combined LPC and DWT based method to extract formant frequencies of speech signal is proposed in [6], in fact, DWT is used for speech signal de-noising before formant estimation with LPC features. In [7], Bensaid also proposed a detection method based on wavelet transform and cepstrum. The method has good results, but it has a

Greater complexity than our method. Extraction and tracking of formant frequencies are important issues in speech processing. The aim of formant tracking is to track the trajectory of the formant frequencies throughout continuous speech signals. A supervised machine learning approach to get these issues using two deep network architectures have been employed: recurrent and convolutional recurrent networks, where, these networks were fed by LPC coefficients and spectrograms, respectively [8].

## IV. FUZZY CLASSIFICATION TECHNIQUES FOR SPEECH PROCESSING APPLICATIONS

The fuzzy logic is a noteworthy technique to countermeasure the analysis and design issues for complex nonlinear systems. In the world of 1-D signals, for a first time fuzzy logic was used in speech signal processing [9]. Various fuzzy systems can be efficiently utilized in order to design a flexible speech recognition system. Several fuzzy techniques to speech and speaker recognition were presented and appraised in [10]. In the following, some fuzzy based techniques proposed in speech recognition systems are reviewed.

*A. Speech and speaker recognition*

Recognition of phonetic units has positive effect in speech enhancement applications. As regard to more energy of vowel phonemes rather than that in consonants, a phonology based fuzzy phoneme recognition system using ZCR and STE features has been proposed in [11]. Jalili et al. [12] suggested a combined fuzzy and Ant Colony (ACO) based speech recognition technique. They has exploited fuzzy concepts to reduce dimensionality as well as to increase recognition rate. After that, the ACO clustering algorithm has been used to classify them to their suitable classes. In additions, the method can distinguish noise and remove it from the original signal [12]. A different feature extraction technique using Diatonic Frequency Cepstral Coefficients (DFCC) resulted from a musical scale called as diatonic scale is introduced in [13]. Here, the scale is according to sound harmonics and it shows nonlinear characteristic of human auditory filter. After feature extraction, a hybrid NF classifier is utilized to distinguish the patterns produced by DFCC algorithm in order to recognize and classify spoken words. The fuzzy logic enhances the recognition rate of the used NN through reducing wrong patterns matching [13].

In [14], a speech recognition design is implemented to control an arm robot. The method uses LPC features and an Adaptive Neuro-Fuzzy Inference System (ANFIS) to distinguish speech signals. Asemi et al. [15] use ANFIS to compare different Multi Views Multi Learners (MVML) and Multi-view Semantic Learning (MVSL) speech recognition methods. By counting the number of identified words, total number of words, and total number of vocabulary and user satisfaction checklist, methods were evaluated in terms of six factors: feedback adequacy, accuracy, user evaluation, error handling, standards and recognition rate. Simulation results showed that the MVML based speech recognition is a better choice in comparison with the MVSL based one. Naini et al. [16] used Formants and formant-gaps (FoGs) features in order to whispered speaker verification. It is shown that, formants with 1st order formant gaps are more important than other formant based features.

*B. Speech segmentation*

Speech segmentation is a challenging subject in speech processing systems. Short term energy, zero-crossing rate and the singularity exponents are the time-domain extracted features which are used in a fuzzy based continuous speech segmentation algorithm [17]. The outputs of this segmentation method are silence, phonemes, or syllables. However, Fuzzy logic is utilized to fuzzify the extracted features into three complementary sets: low, medium, high and to do a matching step using a set of fuzzy rules [17, 18]. A two-stage audio and visual speech processing framework exploiting fuzzy logic is introduced in [19]. The goal of this framework is to employ fuzzy logic to choose the most proper processing strategy to use on a segment by segment basis. In other words, by considering different environmental situations, fuzzy logic let the system to filter the noisy speech, intelligently. To get more knowledge from speech signals, authors of [20] used a voiced/unvoiced detection algorithm based on the multi-scale product analysis by utilizing fuzzy logic. After that, a comb filter is used for the voiced parts and a spectral subtraction is applied on the unvoiced parts of the noisy speech signal. The comb filter is adjusted by an accurate pitch frequency estimation scheme.

C. Speech Gender Recognition

Gender determination is a challenging issue in speaker recognition systems. Different genders have various frequency ranges and fundamental frequency values. Naturally, the fundamental frequency is a good parameter to distinguish male and female voices. Authors of [21] introduced a Mamdani fuzzy logic based gender classification by using two membership functions as frequency and pitch. Based on the input speech, the output is predicted as male, female, and children. In [22], several speech samples of different sexes for the word, 'Close', were used to make a gender identification system based on Neural Network and Fuzzy Logic. After that, the signal energy and power spectrum as well as power spectrum of sound "O" in the word "close" have been utilized to determine an individual speaker. It is obvious, feature selection has been one of the most important factors in gender recognition systems, hence, Meena et al. used Short Time Energy (STE), Zero Crossing Rate (ZCR) and Energy Entropy (EE) as selected features in their fuzzy logic and neural networks based gender classifier [23]. In another research, Jayashankar et al. applied Genetic Algorithm (GA) to choose among these above three features which are responsible for gender classification [24]. Fuzzy logic has shown good efficiency in both recognition and retrieval functions of an individual speaker identification system [22]. Boujnah et al. [25] proposed a hybrid Speaker recognition system composed of formant extraction and Dynamic Time Wrapping method.

*D.  Speech Emotion Recognition*

Emotion recognition is an interdisciplinary area, and it achieves a significant attention of the researchers in the past few years. Automatic recognition of an emotional state intends to attain an interface among the machines and human beings [26]. Emotion recognition from speech signals has several applications like smart healthcare, autonomous voice response systems, assessing situational seriousness by caller affective state analysis in emergency centers, and other smart affective services [27]. It is seen that different emotional states spreads over different frequency bands. Authors of [28] investigated some feature extraction approaches for speech emotion recognition. In [29], pitch and formants are first extracted from the speech signal and then their analysis is carried out to recognize Neutral, Happy and Sad emotional states of the person. The Cepstral analysis method is used for pitch extraction and LPC analysis method is used to extract the formant frequencies. the pitch frequency for both neutral and angry states are quite different, in other words, the mean of pitch frequency for angry emotion is around 336 Hz which is much higher than the mean of neutral emotion which is around 130 Hz. Therefore for identifying the angry and neutral emotions from speech pitch is considered the best indicator. Formants easily identify the happy emotion but cannot detect the neutral and angry emotions [28]. Pitch calculation for speech signal gives better results for Neutral and Angry emotions than the formant frequency estimation method. Happy emotion is recognized satisfactorily using LPC based formant frequency estimation method [29]. Noroozi et al. introduced a vocal-based emotion detection system based on multiclass SVM, decision tree RF, and Adaboost schemes using 14 features including fundamental and formant frequencies, mean autocorrelation, mean noise-to-harmonics ratio, mean harmonics-to-noise ratio and standard deviation. They showed exploiting formants and vocal features improves the emotion determination rates [30]. In [26], using the spectral features and Multiple Support Vector Neural Network (Multi-SVNN) classifier an emotion recognition approach for emotion determination of speech signals is proposed. In [31] the speaker emotion is recognized by exploiting the fuzzy C-means (FCM) clustering algorithm. In fact, classification was performed by applying the FCM algorithm to spectral features which are obtained from MFCC and LPC analysis. A speech emotion recognition scheme based on the features extracted from spectrograms using a deep convolutional neural network (CNN) is presented in [27]. Kamińska et al. proposed an emotion recognition technique using common features such as pitch, formants, energy, MFCC, LPC and Perceptual Linear Prediction (PLP) as well [32].

V. Proposed feature extraction Algorithms

Since the speech fundamental frequency is lower than formant frequencies, two separate algorithms to estimate these frequencies using DWT and filter banks concepts are suggested.

*A.  Discrete wavelet transform and filter banks*

The wavelet transform is a multi-resolution analysis, and it decomposes the input signal by help of two high pass and low pass filters into different frequency bands. In other words, the details (wavelet) and approximate (scale) coefficients will be obtained by applying the high and the low pass filters on the input signal, respectively. In DWT, the scale and position are changed in discrete steps. Usually, the DWT uses scale and position values based on powers of two. The discrete wavelet transform of signal $X(k)$ is defined by the following equation [33]:

$$W(j,k) = \sum_j \sum_k X(k) 2^{-j/2} \psi(2^{-j} n - k) \quad (1)$$

Where, $\Psi(t)$ is the basic or mother wavelet. In DWT, the original signal passes through two filters; a low-pass filter $h(n)$ and a high-pass filter $g(n)$, and emerges as two signals (Fig.2). The successive application of high pass and low pass filtering on the signal can be shown by following formula [33]:

$$Y_{high}[k] = \sum_n x(n) g(2k - n) \quad (2)$$

$$Y_{low}[k] = \sum_n x(n) h(2k - n) \quad (3)$$

Here, $Y_{high}$ and $Y_{low}$ are detail and approximation coefficients which are the outputs of the high pass and low pass filters, respectively.

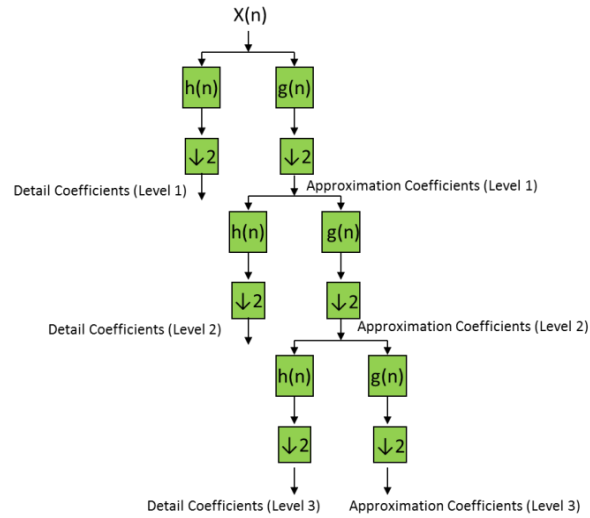

Fig. 2. Wavelet Decomposition Tree Structure [34].

Therefore, at the first stage, the spectrum of signal X (n) is decomposed into two bands of high and low. At the second stage, the resulted low-pass band is separated into a low-pass and a band-pass band. In the other words, at first step, the signal spectrum is divided into two equal bands and in the second step, the given low band itself is divided equally in two bands and this process is repeated. Hence, the input signal is decomposed to the creation logarithmic bands. It is shown that the ratio of a band to the central frequency of the band is always constant, so this framework forms a filter bank.

The results of applying this DWT based filter bank on signal X(.) and thereupon its decomposition into frequency bands are illustrated in Fig.3., specially, the distribution of DWT parameters over frequency bands of a 20Khz sampled speech signal, up to eight scales is given in Table 2.

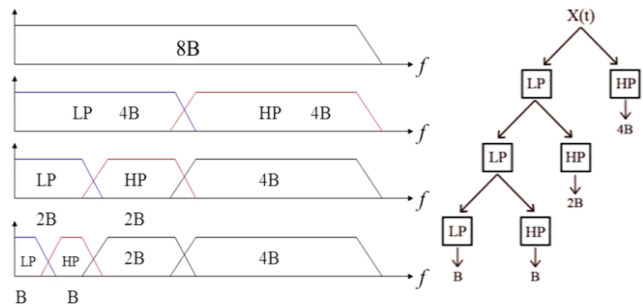

Fig.3. Decomposing a signal spectrum by iterating the filter bank [34].

According to Table.2, the results of lower scales belong to formant frequencies and the resulted data of scales more than 5, contain low frequency components such as pitch frequency.

Table.2. Distribution of Wavelet Parameters over Frequency Bands

| Scale No. | Frequency Band(Hz) |
|---|---|
| 1 | 5000-10000 |
| 2 | 2500-5000 |
| 3 | 1250-2500 |
| 4 | 625-120 |
| 5 | 312.5 -625 |
| 6 | 156.25-312.5 |
| 7 | 78.125-156.25 |
| 8 and above | 0 - 78.125 |

B. *Extraction of formant frequencies*

The range of the pitch frequency is often between 50 and 500 Hz, while the range of formant frequency is between 200 Hz to 8 kHz. Therefore, since the speech pitch frequency is in a lower frequency band compared to the speech formant frequency, the effect of stimulation and vocal cavity can be separated by using a wavelet transform. With each wavelet transformation and the rise of scale, the high-frequency components of the signal are deleted and only low-frequency components of the signal will remain. In other words, the coefficients of approximation and details at high scales are related to low frequency components of the signal. Therefore, in the case of eliminating coefficients in high scales, low frequency components will be eliminated in the analyzed segment. So, we can remove the components related to the pitch frequency from the signal and only the formant frequencies remain in the signal. So the method of work is that after sampling the signal at the appropriate rate, the signal is divided into several segments. In fact, at first, the speech signal is sampled by 20 kHz frequency and then it is partitioned using a Hamming window with a length of 256 samples and jump of 1/4 window length as follows [35]:

$$w(n) = \begin{cases} 0.54 - 0.46 \cos \frac{2\pi n}{N-1} & 0 \leq n < N-1 \\ 0 & else \end{cases} \quad (4)$$

Since, in the Hamming window, the main lobe is large and the lower lobes are smaller, its frequency leakage is lower than the other windows.

By applying DWT, the approximation and details of each partition are calculated up to eighth scale. As regard to Table 2, the obtained wavelet coefficients are zeroed from fifth scale afterwards. Next, the signal is restored by using Inverse Discrete Wavelet Transform (IDWT). The resulted signal includes only formant information components which can be extracted by applying Fourier transform on it. Now, to estimate formant frequencies, it is just enough to compute the smoothed spectrum of the restored signal, and extract its peaks. These peaks are the formant frequencies. The flowchart of the proposed formant extraction method is depicted in Fig.4.

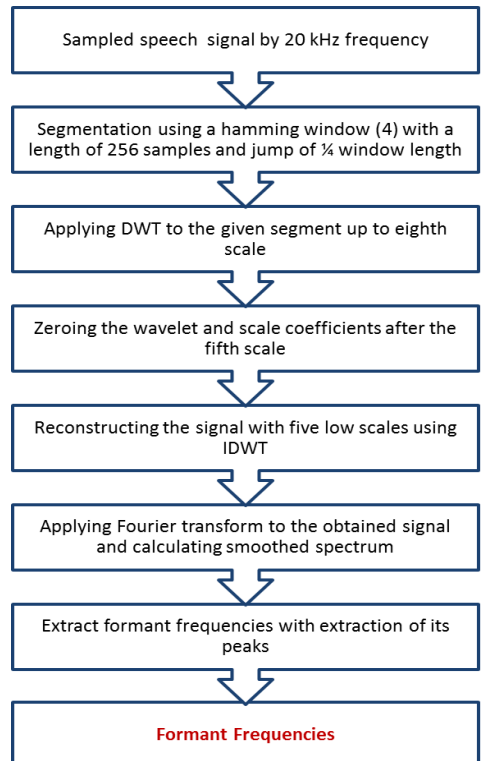

Fig. 4. Block diagram of the proposed algorithm for Detection of Formant Frequencies.

## C. Extraction of the Pitch Frequency

Pitch frequency is lower than the formant frequencies and it is often between 50Hz to 500Hz. Thus, by omitting the frequency components which lie out of this range, the pitch frequency can be extracted easily. It is just enough to eliminate the wavelet coefficients regarded to be out of this range. Therefore, the original signal is sampled by 20 kHz sampling rate and further partitioned using Hamming window with a length of 512 samples. After that, wavelet transform of the obtained partition is calculated up to the ninth scale.

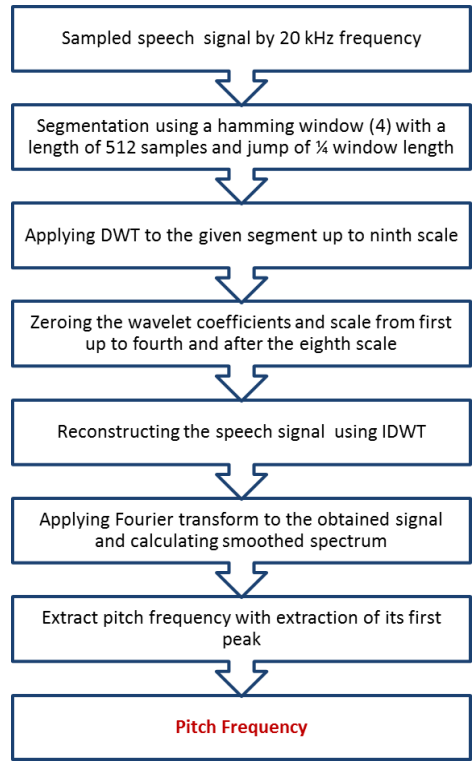

Fig.5. Flowchart of the proposed algorithm for Pitch Frequency Extraction

According to Table.2, wavelet and scale coefficients are zeroed from first up to the fourth scale as well as after the eighth scale. After reconstructing the signal, it includes only pitch information. Next, Fourier transform is applied to the obtained signal and smoothed spectrum is calculated. Therefore, it is possible to extract pitch frequency via extraction of the smoothed spectrum first peak (Fig.5).

## VI. EXPERIMENTAL RESULTS

The proposed methods were applied on different signals and compared with the results obtained from the cepstrum method. The results indicate the efficiency of the wavelet transform over other methods in extraction of pitch and formant frequencies. Fig. 6-a, show a segment of the voiced speech /a/ in time domain and Fig.6-b, illustrate the smoothed spectrum of this signal and the obtained spectrum after removing wavelet coefficients of first up to forth and above the eighth scales as well as the smoothed spectrum of cepstrum.

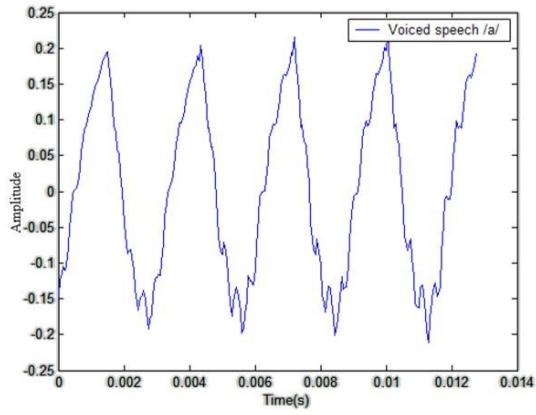

(a)

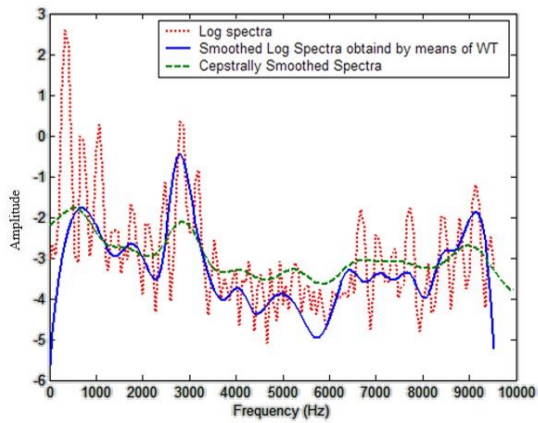

(b)

Fig.6. (a) Original Signal /a/, (b) Comparison between Cepstrum and our formants detection algorithm Spectrums of /a/.

Fig.7-a, shows an unvoiced signal in the time domain while Fig. 7-b, illustrates the obtained spectrum of the original signal and the obtained spectrum after omission of high scales coefficients through wavelet transform as well as the cepstrum. Moreover, Fig. 8-a, and Fig. 9-a, show wave form of two voiced speech signal in time domain while Fig.8-b, and Fig.9-b, illustrate the smoothed spectrum of the signal and the obtained spectrum after removing wavelet coefficients of first up to forth and above the eighth scales as well as the smoothed spectrum of cepstrum.

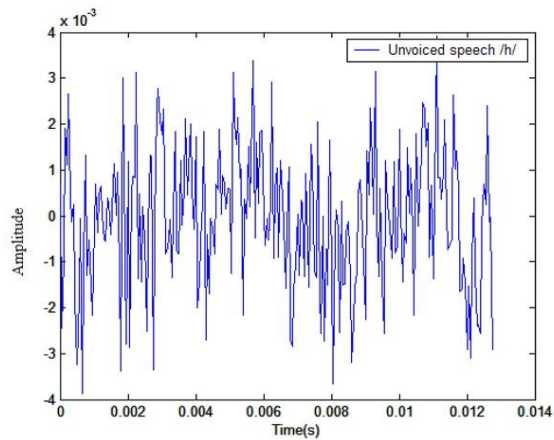

(a)

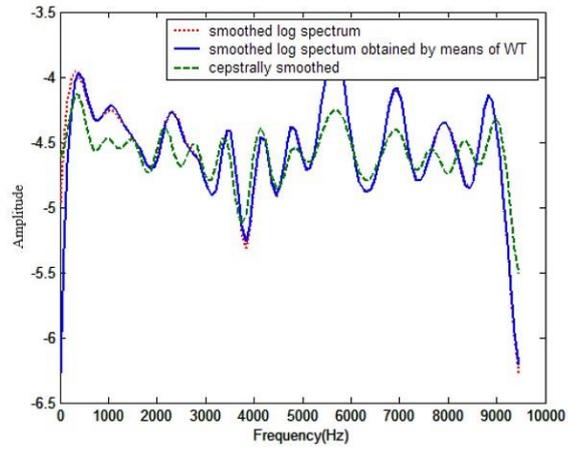

(b)

Fig.7. (a) Original Signal /h/, (b) Comparison between Cepstrum and our formants detection algorithm Spectrums of /h/.

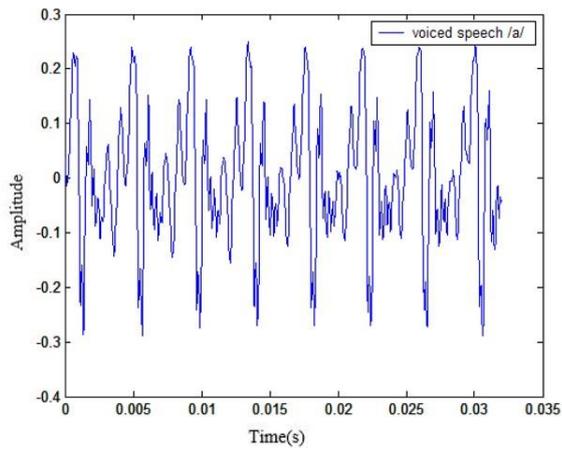

(a)

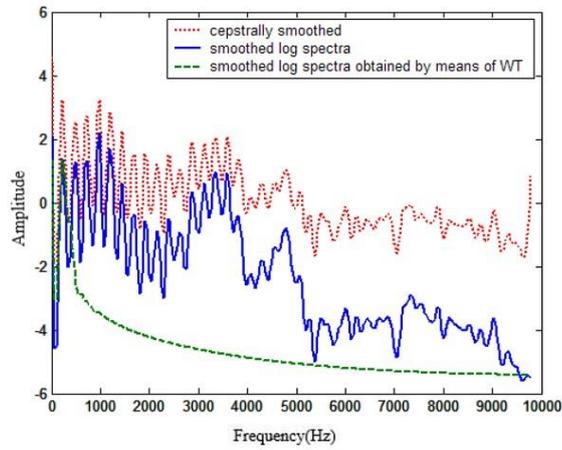

(b)

Fig. 8. (a) Original Signal /a/, (b) Comparison between Cepstrum and our Pitch detection algorithm Spectrums of /a/.

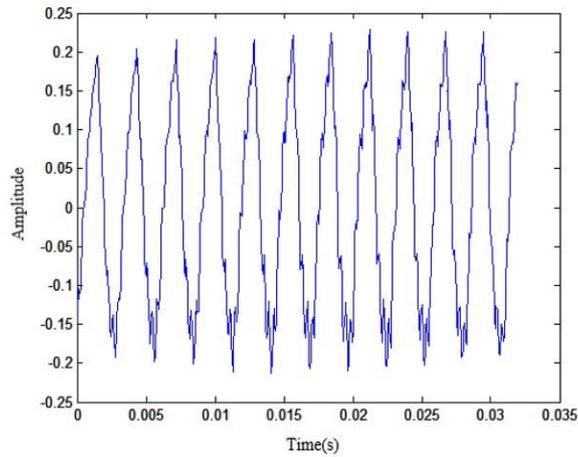

(a)

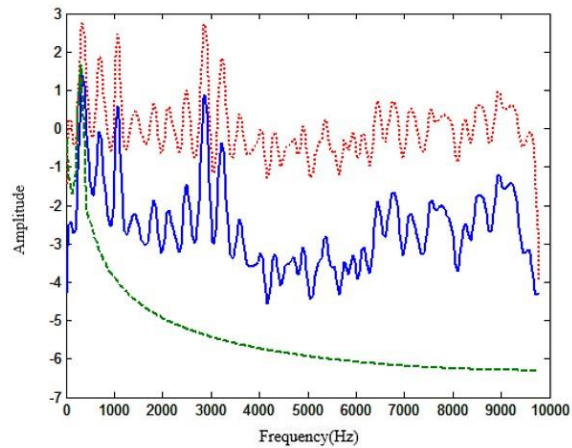

(b)

Fig.9. (a) An original voiced Signal, (b) Comparison between Cepstrum and our pitch detection algorithm Spectrum.

VII. CONCLUSION

This article proposes two DWT based algorithms to extract pitch and formant frequencies. These methods were examined by simulating different samples. By comparing the methods presented for pitch extraction on several speech signals, it was found that the wavelet transform has a good accuracy compared to the cepstrum method and in has no sensitivity to noise. However, noise can be eliminated from signals using the wavelet transform. In addition, it's easy to distinguish the peaks in some scales and calculate the pitch period. Furthermore, regarding different wavelets, Daubishes family, especially Daubishes 20 proved to be more appropriate than the others in feature extraction of speech signals. As a future work, we are going to present a new fuzzy classifier regard to the extracted features.